\documentstyle[preprint,prb,aps,graphicx]{revtex}

\begin{document}

\title{A Fast and Compact Quantum Random Number
Generator}

    \author{Thomas Jennewein, Ulrich Achleitner\dag, Gregor Weihs, \\
    Harald Weinfurter\ddag, and Anton Zeilinger}

    \address{Institut f\"{u}r Experimentalphysik, Universit\"{a}t Wien, \\
    Boltzmanngasse 5, A--1090  Wien, Austria \\
    \dag Institut f\"{u}r Experimentelle Anaesthesie,
    Universit\"{a}tsklinik f\"{u}r Anaesthesie und Intensivmedizin,
    Anichstra{\ss}e 35, A--6020  Innsbruck, Austria \\
    \ddag Sektion Physik, Ludwig-Maximilians-Universit\"{a}t Muenchen\\
    Schellingstr. 4/III  D-80799 M\"{u}nchen, Germany}

\date{\today}

\maketitle

\thispagestyle{empty}



\begin{abstract}
We present the realization of a physical quantum random number
generator based on the process of  splitting a beam of photons on
a beam splitter, a quantum mechanical source of true randomness.
By utilizing either a beam splitter or a polarizing beam splitter,
single photon detectors and high speed electronics the presented
devices are capable of generating a binary random signal with an
autocorrelation time of $11.8$~ns and a continuous stream of
random numbers at a rate of 1~Mbit/s. The randomness of the
generated signals and numbers is shown by running a series of
tests upon data samples. The devices described in this paper are
built into compact housings and are simple to operate.
\end{abstract}



\newpage

\section{Introduction}
Random numbers are a vital ingredient in many applications ranging,
to name some examples, from computational methods such as Monte Carlo
simulations and programming \cite{AGA89}, over to the large field of
cryptography for generating of crypto code or masking messages, as
far as to commercial applications like lottery games and slot
machines \cite{MA93,MACD,Maurer90,Ellison}. Recently the range of
applications requiring random numbers was extended with the
development of quantum cryptography and quantum information
processing \cite{BB92}. Yet a novelty is the application for which
the random number generator presented in this paper was developed
for: an experiment regarding the entanglement of two particles, a
fundamental concept within quantum theory \cite{gregor}. Firstly this
experiment demanded the generation of random signals with an
autocorrelation time of $<100$~ns. Secondly, for the clarity of the
experimental results, it was necessary that true (objective)
randomness was implemented.

The range of applications using random numbers has lead both to
the development of various random number generators as well as to
the means for testing the randomness of their output. Generally
there are two approaches of random number generation, the pseudo
random generators which rely on algorithms that are implemented on
a computing device, and the physical random generators that
measure some physical observable expected to  behave randomly.

Pseudo random generators are based on algorithms or even a
combination of algorithms and have been highly refined in terms of
repetition periods\cite{KU94} ($2^{800}$) and robustness against
tests for randomness\cite{MA93}. But the inherent algorithmic
evolution of pseudo random generators is an essential problem in
applications requiring unpredictable numbers as the unguessability
of the numbers relies on the randomness of the seeding of the
internal state. Dependent on the intended application this can be
a drawback. The requirements of our specific implementation were
even such, that the use of a pseudo random number generator was in
itself already ruled out by its deterministic nature.

Physical random generators use the randomness or noise of a physical
observable, such as noise of electronic devices, signals of
microphones, etc. \cite{Ellison}. Many such physical sources utilize
the behavior of a very large and complex physical systems which have
a chaotic, yet at least in principle deterministic, evolution in
time. Due to the many unknown parameters of large systems their
behavior is taken for true randomness. Still, purely classical
systems have a deterministic nature over relevant time scales, and
external influences into the random generator may remain hidden.

Current theory implies that the only way to realize a clear and
understandable physical source of randomness is the use of elementary
quantum mechanical decisions, since in the general understanding the
occurrence of each individual result of such a quantum mechanical
decision is objectively random (unguessable, unknowable). There
exists a range of such elementary decisions which are suitable
candidates for a source of randomness. The most obvious process is
the decay of radioactive nucleus ($^{85}\mathrm{Kr}$,
$^{60}\mathrm{Co}$) which has already been used \cite{IS56,WA96}.
However, the handling of radioactive substances demands extra
precautions, especially at the radioactivity required by the
switching rates of our envisaged random signals. Optical processes
suitable as a source of randomness are the splitting of single photon
beams \cite{AC97}, the polarization measurement of single photons,
the spatial distribution of laser speckles \cite{MA91,MA86} or the
light-dark periods of a single trapped ion's resonance fluorescence
signal \cite{IT87,SA86}. But only the first two of the mentioned
optical processes are fast enough and, in addition, do not require an
overwhelming technical effort in their realization. Thus we developed
a physical quantum mechanical random generator based on the splitting
of a beam of photons with an optical 50:50 beam splitter or by
measuring the polarization of single photons with a polarizing beam
splitter.\cite{AC97}

\section{Theory of Operation}

\begin{figure}

\includegraphics{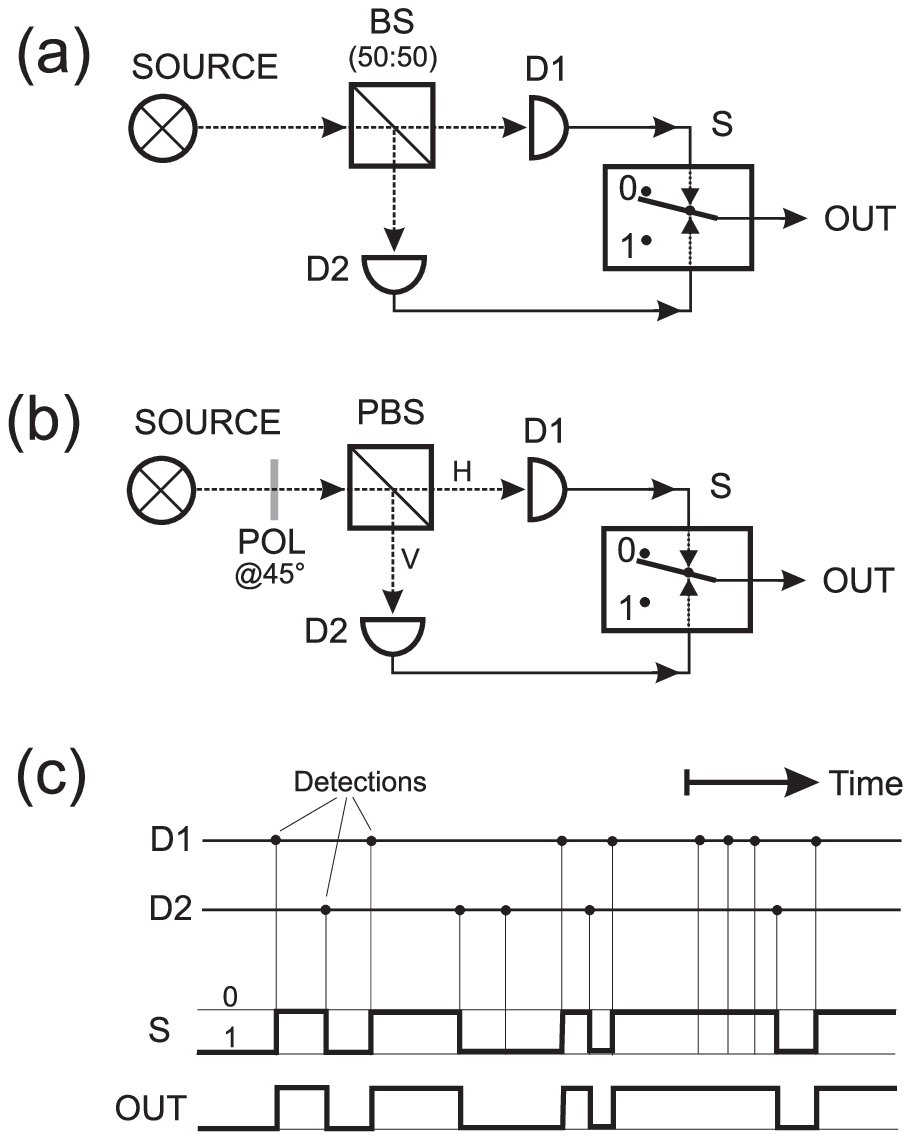}

\vspace{0.7cm}

\caption{The source of randomness within our device is the
splitting of a weak light beam. This is realized with a 50:50
optical beam splitter BS (a) or a polarizing beam splitter PBS
where the incoming light is polarized with POL at $45^{\circ}$
with respect to the PBS (b). The photon detections of the
detectors D1 and D2 in the two output paths toggle the switch S
between its two states (c). This produces a randomly alternating
binary signal OUT.}

\label{prinzip}

\end{figure}

The principle of operation of the random generator is shown in
Figure~\ref{prinzip}.  For the case of the $50:50$ beam splitter (BS)
(Figure~\ref{prinzip}(a)), each individual photon coming from the
light source and traveling through the beam splitter has, for itself,
equal probability to be found in either output of the beam splitter.
If a polarizing beam splitter (PBS) is used
(Figure~\ref{prinzip}(b)), then each individual photon polarized at
$45^{\circ}$ has equal probability to be found in the H (horizontal)
polarization or V (vertical) polarization output of the polarizer.
Anyhow, quantum theory predicts for both cases that the individual
``decisions'' are truly random and independent of each other. In our
devices this feature is implemented by detecting the photons in the
two output beams with single photon detectors and combining the
detection pulse in a toggle switch (S), which has two states, 0 and
1. If detector D1 fires, then the switch is flipped to state 0 and
left in this state until a detection event in detector D2 occurs,
leaving the switch in state 1, until a event in detector D1 happens,
and S is set to state 0. (Figure~\ref{prinzip}(c)). In the case that
several detections occur  in a row in the same detector, then only
the first detection will toggle the switch S into the corresponding
state, and the following detections leave the switch unaltered.
Consequently, the toggling of the switch between its two states
constitutes a binary random signal with the randomness lying in the
times of the transitions between the two states. In order to avoid
any effects of photon statistic of the source or optical interference
onto the behaviour of the random generator the light source should be
set to produce $\ll 1$ photon per coherence time.

\section{Realization of the Device}
\begin{figure}

\includegraphics[width=14cm]{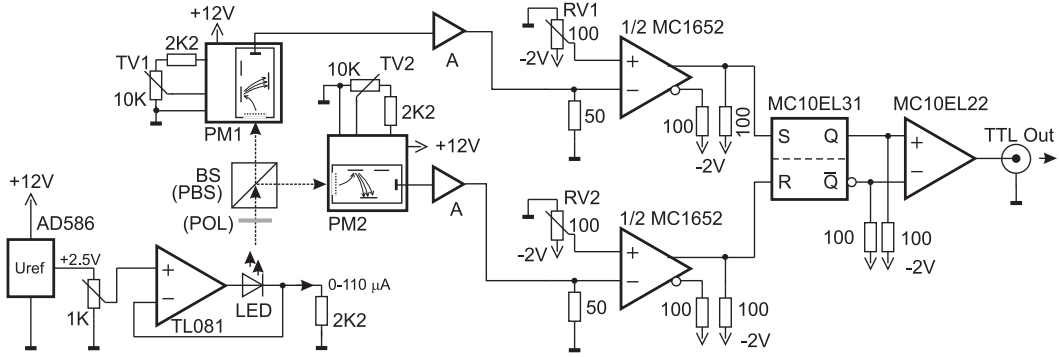}

\vspace{0.7cm}

\caption{Circuit diagram of the physical quantum random generator.
To the left is the light source (LED) and the configuration of the
beam splitter (BS/PBS) and the two photo multipliers (PM1, PM2).
The detection pulses of the PMs are turned into standard logic
pulses with discriminators (MC1652) and combined in the
RS--flip-flop (MC10EL31) to generate the random signal.}

\label{schaltbild}

\end{figure}

Figure~\ref{schaltbild} shows the circuit diagram of the physical
quantum random generator. The light source is a red light emitting
diode (LED) driven by an adjustable current source (AD586 and TL081)
with maximally $110\ \mu\mathrm{A}$. Due to the very short coherence
length of this kind of source ($<1$~ps) it can be ascertained, that
most of the time there are no photons present within the coherence
time of the source, thus eliminating effects of source photon
statistics or optical interference. The light emerging from the LED
is guided through a piece of pipe to the beam splitter, which can be
either a 50:50 beam splitter or a polarizing beam splitter. In the
latter case the photons are polarized beforehand with polarization
foil (POL) at $45^\circ$ with respect to the axis of the dual channel
polarization analyzer (PBS). The photons in the two output beams are
detected with fast photo multipliers\cite{pmt} (PM1, PM2). The PMs
are enclosed modules which contain all necessary electronics as well
as a generator for the tube voltage, and thus only require a $+12$~V
supply. The tube voltages can be adjusted with potentiometers (TV1,
TV2) for
 optimal detection pulse rates and pulse amplitudes . The
output signals  are amplified in two Becker\&Hickl amplifier
modules (A) and transmitted to the signal electronics which is
realized in emitter-coupled-logic (ECL). The detector pulses are
converted into ECL signals by two comparators (MC1652) in
reference to adjustable threshold voltages set by potentiometers,
(RV1, RV2). The actual synthesis of the random signal is done
within a RS--flip-flop (MC10EL31) as PM1 triggers the {\bf
S}--input and PM2 triggers the {\bf R}--input of the flip-flop.
The output of this flip-flop toggles between the high and low
state dependent upon whether the last detection occurred in PM1 or
PM2. Finally the random signal is converted from ECL to TTL logic
levels (MC10EL22) for further usage.

\begin{figure}

\includegraphics{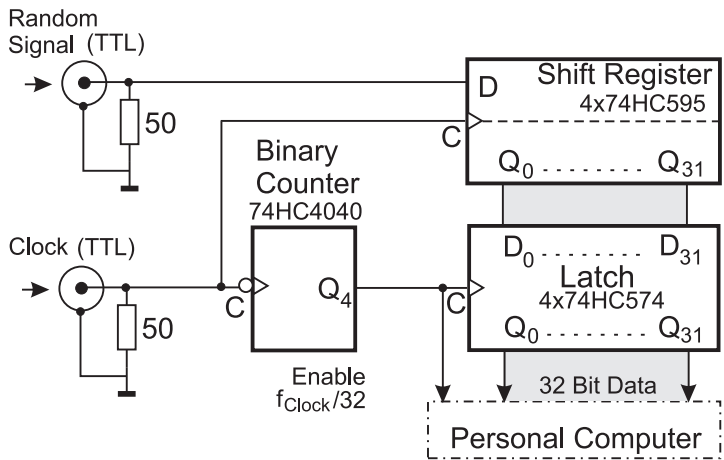}

\vspace{0.7cm}

\caption{Schematic diagram of the circuit for transferring random
numbers to a personal computer at a constant bit rate.}

\label{sampling}

\end{figure}

In order to generate random numbers on a personal computer the
signal from the random generator is sampled periodically and
accumulated in a 32~bit wide shift register
(Figure~\ref{sampling}). Every 32 clock cycles the contents of the
shift register are transferred in parallel to a personal computer
via a fast digital I/O board. In this way a continuous stream of
random numbers is transferred to a personal computer.

\section{Testing the Randomness of the Device}
Up to now, no general definition of randomness exists and discussions
still go on. Two reasonable and widely accepted conditions for the
randomness of any binary sequence is its being ``chaotic'' and
``typical''. The first of these concepts was introduced by Kolmogorov
and deals with the algorithmic complexity of the sequence, while the
second originates from Martin-L\"{o}v and says that no particular random
sequence must have any features that make it distinguishable from all
random sequences \cite{CO91b,US90}. With pseudorandom generators it
is always possible to predict all of their properties by more or less
mathematical effort, due to the fact of knowing their algorithm. Thus
one may easily reject their randomness from a rigorous point of view.
In  contrast, the mostly desired feature of a true random generator,
its ``truth'', bears the principal impossibility of ever describing
such a generator completely and proving its randomness beyond any
doubt. This could only be done by recording its random sequence for
an infinite time. One is obviously limited experimentally to finite
samples taken out of the infinite random sequence. There are lots of
empirical tests, mostly developed in connection with certain Monte
Carlo simulation problems, for testing the randomness of such finite
samples \cite{MA93,LE92}. The more tests one sample passes, the
higher we estimate its randomness. We estimate a test for randomness
the better, the smaller or more hidden the regularities may be that
it can detect \cite{AC97,VA95}.

As the range of tests for the randomness of a sequence is almost
unlimited we must find tests which can serve as an appropriate
measure of randomness according to the specific requirements of
our application. Since the experiment that our random generators
are designed for demands random signals at a high rate, we focus
on the time the random generators take to establish a random state
of its signal starting from a point in time where the output state
and the internal state of the generator may be known.

We will briefly describe the relatively intuitive tests that will
be applied to data samples taken from the random generator, which
we consider to be sufficient in qualifying the device for its use
in the experiment.

\begin{enumerate}

\item {\bf Autocorrelation Time of the Signal:} For a binary sequence as
produced by our random generator the autocorrelation function
exhibits an exponential decay of the form:\cite{mandel}
\begin{equation}
A(\tau)=A_0\mathrm{e}^{-2R|\tau|},
\end{equation}
where $R$ is the average toggle rate of the signal, $A_0$ is the
normalization constant and $\tau$ is the delay time. Per
definition the autocorrelation time is given by
\begin{equation}
\tau_{ac}=\frac{1}{2R}. \label{autocorrelationtime}
\end{equation}
The autocorrelation function is a measure for the average
correlation between the signal at a time $t$ and later time
$t+\tau$.

\item {\bf Internal Delay within the Device:} The internal delay time
within the device between the emission of a photon and its effect on
the output signal. This internal delay time is the minimal time the
generator needs to establish a truly random state of its output.

\item {\bf Equidistribution of the Signal:} This is the most obvious and
simple test of randomness of our device, as for random generator
the occurrence of each event must be equally probable. Yet, by
itself the equidistribution is not a criterion for the randomness
of a sequence.

\item {\bf Distribution of Time Intervals between Events:} The transitions
of the signal generated by our system are independent of any
preceding events and signals within the device. For such a
Poissonian process the time intervals between successive events
are distributed exponentially in the following way:
\begin{equation}
p(T)=p_0\mathrm{e}^{-T/T_0},
\end{equation}
where $p(T)$ is the probability of a time interval $T$ between two
events, $T_0=1/R$ is the mean time interval and is the reciprocal
value of the average toggle rate $R$ defined earlier and $p_0$ is
the normalization constant. The evaluation of $p(T)$ for a data
sample taken from our generator shows directly for which time
intervals the independence between events is ascertained and for
which time intervals the signal is dominated by bandwidth limits
or other deficiencies within the system.

\item {\bf Further Illustrative Tests of Randomness:}
These statistical tests will be applied to samples of random
numbers produced by the random generator in order to illustrate
the functionality of the device. For the application our random
generators are designed for these statistical measures are not as
important as the tests described above, and the tests proposed
here represent just a tiny selection of possible test. Yet, these
tests allow a cautious comparison of random numbers produced with
our device with random numbers taken from other sources. The code
for the evaluation evaluation of these tests was developed in
\cite{AC97}.

\begin{enumerate}

\item {\bf Equidistribution and Entropy of $n$--Bit Blocks:} Provided
that the sample data set is sufficiently long, all possible
$n$--bit blocks (where $n$ is the length of the block) should
appear with equal probability within the data set. A direct, but
insufficient, way of determining the equidistribution of a data
set is to evaluate the mean value of all $n$--bit blocks, which
should be $(2^n-1)/2$. This will give the same result for any
symmetric distribution. The distribution of $n$--bit blocks of a
data set corresponds to the entropy, a value which is often used
in the context of random number analysis. The entropy is defined
as:
\begin{equation}
H_n=-\sum_np_i\log_2p_i
\end{equation}
and is expressed in units of bits. $p_i$ is the empirically
determined probability for finding the $i$--th block. For a set of
random numbers a block of the length $n$ should produce $n$ bits
of entropy. In the case of bytes, which are blocks of 8 bits, the
entropy of these blocks should be 8 bits.

\item {\bf Blocks of $n$ Zeros or Ones:} Another test for the
randomness of a set of bits is the counting of blocks of
consecutive zeros or ones. Each bit is equally likely a zero as a
one, therefore the probability of finding blocks of $n$
concatenated zeros or ones should be proportional to a $2^{-n}$
function.

\item {\bf Monte-Carlo estimation of $\pi$:} A pretty way of
demonstrating the quality of a set of random numbers produced by a
random generator is a simple Monte Carlo estimation of $\pi$. The
idea is to map the numbers onto points on a square with a quarter
circle fitted into the square and count the points which lay
within the quarter circle. The ratio of the number of points lying
in the circle and the total number of points is an estimation of
$\pi$.

\end{enumerate}

\end{enumerate}

\section{Operation of the Device}
Two random generators were each built into single width NIM-module
(dimensions: $25*19*3\mathrm{cm}^3$) in order to match our existing
equipment. The optical beam splitter, the two photo multipliers and
the pulse amplifiers are mounted on a base-plate within the modules
and the electronics is realized on printed circuit boards. The random
generator modules require only a standard voltage supply of $\pm6$~V
and $+12$~V. The random signal generators were configured either with
a 50:50 beam splitter or with a polarizing beam splitter as the
source of randomness. In both cases they performed equally well. Yet,
the polarization measurement of the photons offers the advantage of
adjusting the division ratio of the photons in the two beams by
slightly rotating the polarization foil sitting just in front of the
beam splitter. The results presented here were all obtained from a
random signal generator configured with the polarizing beam splitter.

After warm up the devices require a little adjustment for maximum
average toggle rate and equidistribution of the output signal. The
average toggle rate of the random signals is checked with a
counter and the equidistribution is checked by sampling the signal
a couple of thousand times and counting the occurrences of zeros
and ones. These measures are both optimized by trimming the
reference voltage of the discriminators (RV1, RV2) and by
adjusting the tube voltages of the photo multipliers (TV1, TV2).
The  maximum average toggle rate of the random signals at the
output of the random number generators is $34.8$~MHz. Once the
devices are set up in this way they run stably for many hours.

Typically the PMs produce output pulses with an amplitude of
maximally $50$~mV at a width of $2$~ns. The rise and fall time of
the signals produced by the random number generators is $3.3$~ns.
As it turns out, this limit is set by the output driver stage of
the electronics. The transition times of the internal ECL signals
was measured to be less then $1$~ns, which is in accordance with
the specifications of this ECL logic.

\section{Performance of the Device}
The time delay between the emission  of a photon from the light
source and its effect on the output signal after running through the
detectors and the electronics was measured by using a pulsed light
source instead of the continuous LED and observing the electronic
signals within the generator on an oscilloscope. The total time delay
between a light pulse and its effect on the output was $75$~ns, and
consists of $20$~ns time delay in the light source, light path and
the detection, $20$~ns time delay in the amplifiers and cables and
$35$~ns time delay in the main electronics.

\begin{figure}

\includegraphics{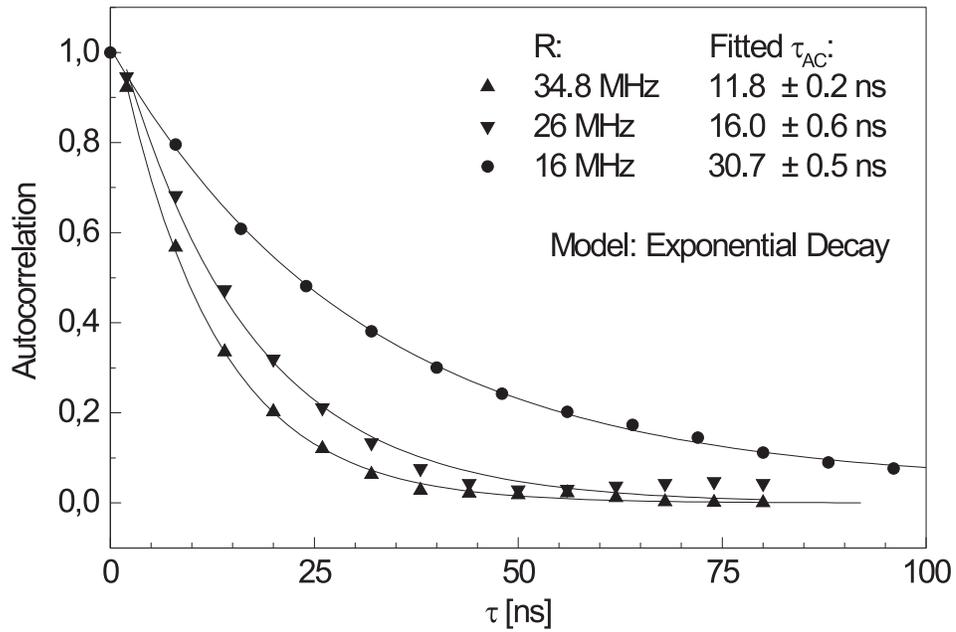}

\vspace{0.7cm}

\caption{The autocorrelation functions computed from traces of the
random signal with different average signal toggle rates. The
given autocorrelation times are obtained with an exponential decay
fit.}

\label{autocorrelation}
\end{figure}

In order to evaluate the autocorrelation time of the signal
produced by the device, signal traces consisting of 15000 points
were recorded on a digital storage oscilloscope for three
different average toggle rates. The sampling rate  of the
oscilloscope  was $500$~MS/s for the toggle rates of $34.8$~MHz
and $26$~MHz and $250$~MS/s for the toggle rate of $16$~MHz. The
autocorrelation function for each trace was evaluated on a
personal computer and the autocorrelation time $\tau_{AC}$ was
extracted by fitting an exponential decay model to these
functions. (Figure~\ref{autocorrelation}) The resulting
autocorrelation times are $11.8 \pm 0.2$~ns ($14.4$~ns) for the
$34.8$~MHz signal, $16.0 \pm 0.6$~ns ($19.2$~ns) for the $26$~MHz
signal and $30.7 \pm 0.5$~ns ($31.3$~ns) for the $16$~MHz signal
which are comparable with the autocorrelation times calculated
with expression~(\ref{autocorrelationtime}) from the average
toggle rate $R$, given in parenthesis.

\begin{figure}

\includegraphics{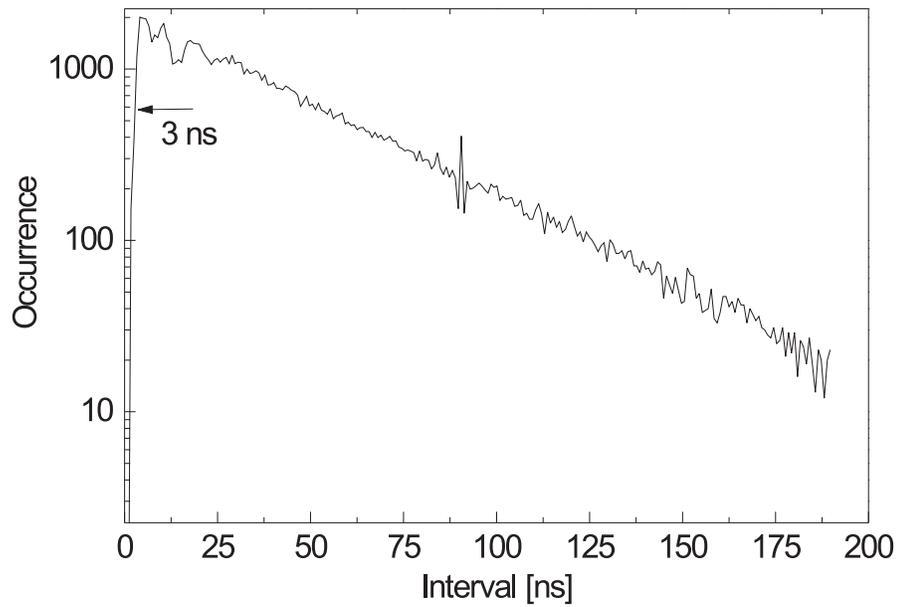}

\vspace{0.7cm}

\caption{Distribution of $10^6$ time intervals between successive
transitions of the random signal with an average toggle rate of
$26$~MHz. This distribution follows an exponential decay function
for times $>35$~ns. (The spike at $96$~ns is from the counter
itself.) }

\label{deltat}
\end{figure}

The time difference between successive toggle events of the random
signal is measured with a time interval counter. The start input
of the counter is triggered by the positive transition, and the
stop input is triggered by the negative transition of the signal.
Figure~\ref{deltat} shows the distribution of $10^6$ time
intervals for a random signal with an average signal toggle rate
of $26$~MHz. For times $<3$~ns the transition time of the
electronics between the two logical states becomes evident as a
cutoff in the distribution. For intervals of up to $35$~ns some
wiggles of the distribution are apparent. This is most likely due
to ringing of the signals on the transmission line. For times
$>35$~ns the distribution approaches an exponential decay
function. The spike at $96$~ns was identified as an artifact of
the counter due to an internal time marker.

As described earlier, our device can produce random numbers by
periodically sampling the signal and cyclically transferring the
data to a personal computer. Our personal computer (Pentium
processor, $120$~MHz, $144$~MB RAM, running LabView on Win95)
manages to register sets of random numbers up to a size of
$15$~MByte in a single run at a maximum sample rate  of $30$~MHz.
In order to obtain independent and evenly distributed random
numbers, the sampling period must be well above the
autocorrelation time of the random signal. We observed that for a
signal autocorrelation time of roughly $20$~ns a sampling rate of
$1$~MHz suffices for obtaining ``good'' random numbers.

\begin{figure}

\includegraphics{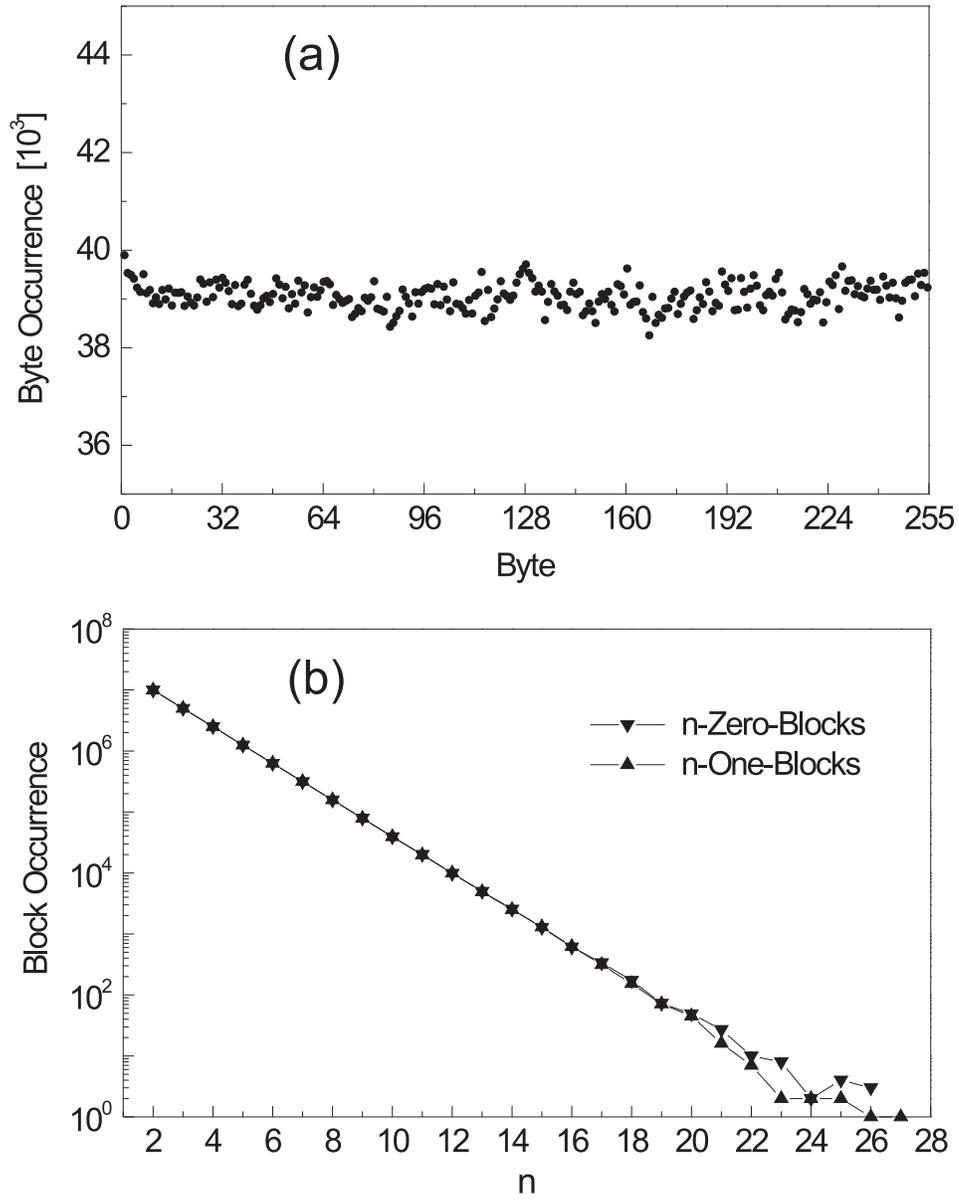}

\vspace{0.7cm}

\caption{Diagram (a): distribution of bytes within $80\cdot10^6$
random bits produced with the random number generator. Diagram
(b): the occurrence of blocks of concatenated zeros and ones
within the same set of random numbers.}

\label{blocks}

\end{figure}

All data samples used for the following evaluations consisted of
$80\cdot10^6$~bits produced in continuous runs with a $1$~MHz bit
sampling frequency. Figure~\ref{blocks}(a) depicts the
distribution of blocks with $8$~bit length within a data sample.
This distribution approaches an even distribution, but still shows
some non-statistical deviations, such as a peak at in the center
and some symmetric deviations. Possibly this is due to a yet to
high sampling rate and a slight misadjustment of the generator.
The distribution of blocks of $n$ concatenated zeros and ones
within a sample should be proportional to a $2^{-n}$--function.
(Figure~\ref{blocks}(b)) The slopes of the logarithmically scaled
distributions are measured with a linear fit and are $-0,29725 \pm
0,00121$ for the $n$--zero blocks and  $-0.30299 \pm 0.00138$ in
the case of the $n$--one blocks. Ideally, the slopes should both
be equal to $-\log(2)= -0,30103$. The deviation can be understood
as a consequence of minor differences in the probabilities of
finding a zero or a one at the output of the generator, again due
to misadjustment of the generator.

The mean value of $8$-bit blocks, the entropy for $8$-bit blocks
and the Monte Carlo estimation of $\pi$ are evaluated for a data
sample produced by our random generator and compared to data
samples taken from the Marsaglia CD--ROM \cite{MACD} and a sample
data set built with the Turbo C++ random function \cite{AC97}.
(Table~\ref{table})

The results in Table~\ref{table} are in favor of our device but
the numbers must be treated with caution, as they represent only a
comparison of single samples which may not be representative.

\section{Discussion and Outlook}
The experimental results presented in the chapter above gives
strong support to the expectation, that our physical quantum
random generator is capable of producing a random binary sequence
with an autocorrelation time of $12$~ns and internal delay time of
$75$~ns. This underlines the suitability of these devices for
their use in our specific experiment demanding random signal
generators with a time for establishing a random output state to
be less than $<100$~ns, which is easily achieved by the physical
quantum random generators presented in this paper.

The high speed of our random generators is made possible by the
implementation of state of the art technology using fast single
photon detectors as well as high speed electronics. Moreover, the
collection of tests applied to the signals and random numbers
produced with our quantum random generator demonstrate the quality
of randomness that is obtained by using a fundamental quantum
mechanical decision as a source of randomness.

Some methods for enhancing the performance, be it in terms of signal
equidistribution and/or autocorrelation time, can be foreseen. For
instance, a different method for generating the random signal would
be that each of the PM's toggles a $\frac{1}{2}$-divider which
results in evenly distributed signals. These signals could be
combined in an XOR-gate in order to utilize the quantum randomness of
the polarization analyzer, but fully keeping the equidistribution of
the signal. A reduction of the signal autocorrelation time is
possible by optimizing the signal electronics for speed (e.g. using
ECL signals throughout the design). Further, it is simple to
parallelize several such random generators within one single device,
as there is no crosstalk between the subunits, since the elementary
quantum mechanical processes are completely independent and
undetermined. Hence designing a physical quantum random number
generator capable of producing true random numbers at rates
$>100$~MBit/s or even above 1~GBit/s is a feasible task \cite{LA93}.

We believe that random generators designed around elementary
quantum mechanical processes will eventually find many
applications for the production of random signals and numbers,
since the source of randomness is clear and the devices operate in
a straightforward fashion.


\section{Acknowledgement}
This work was supported by the Austrian Science Foundation (FWF),
project S6502, by the U.S. NSF grant no. PHY 97-22614, and by the
APART program of the Austrian Academy of Sciences.



\newpage

\addtocounter{page}{2}

\newpage

\newpage

\begin{table}

\begin{tabular}{c|c|c|c|c}
           & QRNG       & Bits52     & Canada     & C++  \\
           \hline \hline
  Mean    & $127.50$   & $127.50$   & $126.58$   & $127.22$\\
  Entropy & $7.999965$ & $7.999982$ & $7.997665$ & $7.81118$\\
  $\pi$   & $3.14017$  & $3.14367$  & $3.15789$  & $3.15761$

\end{tabular}

\vspace{0.7cm}

\caption{Evaluation of tests of randomness for data samples taken
from a selection of sources. QRNG: data set generated with our
physical quantum random generator, Bits52: taken from the
Marsaglia CD-ROM \protect\cite{MACD}, data set generated with a
combination of pseudo random generators, Canada: taken from the
Marsaglia CD-ROM, data set generated with a commercial physical
random generator, C++: data set generated with the pseudo random
generator within Turbo C++ (Borland Inc.) \protect\cite{AC97}.}

\label{table}

\end{table}

\end{document}